\documentclass[prl,reprint,superscriptaddress]{revtex4-1}
\usepackage{amsmath}
\usepackage{amssymb}
\usepackage{graphicx}
\usepackage[caption=false]{subfig}
\usepackage{enumitem}
\usepackage{color}
\usepackage[percent]{overpic}
\usepackage{bm}
\usepackage{rotating}
\usepackage{hyperref}
\usepackage{pbox}
\usepackage{array}
\usepackage[normalem]{ulem} % new added \sout{} to strike out
\usepackage[normalem]{ulem}

\newcommand{\ket}[1]{{\left| {#1} \right>}}

\newcommand{\ii}{\mathrm{i}}

\DeclareMathOperator{\tr}{Tr}
\begin{document}

\title{
%Stabilizing quantum dynamics through uncertainty in control parameters
Stabilizing quantum dynamics through coupling to a quantized environment}

\author{Meenu Kumari}
\affiliation{Institute for Quantum Computing, University of Waterloo, Waterloo, Ontario, N2L 3G1, Canada}
\affiliation{Department of Physics and Astronomy, University of Waterloo, Waterloo, Ontario, N2L 3G1, Canada}
\author{Eduardo Mart\'{i}n-Mart\'{i}nez}
\affiliation{Department of Applied Mathematics, University of Waterloo, Waterloo, Ontario, N2L 3G1, Canada}
\affiliation{Institute for Quantum Computing, University of Waterloo, Waterloo, Ontario, N2L 3G1, Canada}
\affiliation{Perimeter Institute for Theoretical Physics, 31 Caroline St N, Waterloo, Ontario, N2L 2Y5, Canada}

\author{Achim Kempf}
\affiliation{Department of Applied Mathematics, University of Waterloo, Waterloo, Ontario, N2L 3G1, Canada}
\affiliation{Institute for Quantum Computing, University of Waterloo, Waterloo, Ontario, N2L 3G1, Canada}
\affiliation{Perimeter Institute for Theoretical Physics, 31 Caroline St N, Waterloo, Ontario, N2L 2Y5, Canada}

\author{Shohini Ghose}
\affiliation{Department of Physics and Computer Science, Wilfrid Laurier University, Waterloo, Ontario, N2L 3C5, Canada}
\affiliation{Institute for Quantum Computing, University of Waterloo, Waterloo, Ontario, N2L 3G1, Canada}
\affiliation{Perimeter Institute for Theoretical Physics, 31 Caroline St N, Waterloo, Ontario, N2L 2Y5, Canada}

%\date{\today}

\begin{abstract}
We show that introducing a small uncertainty in the parameters of quantum systems can make the dynamics of these systems robust against perturbations. 
Concretely, for the case where a system is subject to perturbations due to an environment, we derive a lower bound on the fidelity decay, which increases with increasing uncertainty in the state of the environment. Remarkably, this robustness in fidelity can be achieved even in fragile chaotic systems. We show that non-Markovianity is necessary for attaining robustness in the fidelity.
\end{abstract}

\maketitle

The concept of fidelity decay was originally introduced by Asher Peres~\cite{peres1984stability} as an indicator of chaos in quantum systems and as a tool to understand irreversibility in quantum physics. It measures the overlap of two states evolving under slightly different Hamiltonians, starting from the same initial state. Over the last decade, fidelity decay has become a subject of immense interest in various fields including quantum information (as a measure of stability of quantum motion against perturbations), statistical physics (as Loschmidt echo) and quantum chaos \cite{Gorin2006}. Past studies of fidelity decay have focused on different aspects such as the connection between fidelity decay and the classical notion of Lyapunov exponents \cite{Jalabert2001}, characterization of quantum chaos \cite{emerson2002,gregor2004}, the effect of different kinds of perturbations and in different time regimes \cite{Jacquod2001, Cucchietti2002, benenti2002quantum,Stockmann2008}, conditions for anomalous slow decay or freeze of fidelity \cite{prosen2005fidelity,gorin2006anomalous,Wu2009}, perturbation independent fidelity decay \cite{Jalabert2001}, and connections to decoherence \cite{prosen2003theory,cucchietti2003decoherence,PROSEN2005,casabone2010discrepancies}. Many of these studies have focused on a particular type of perturbation - changes in a control parameter of the system. Examples include variation of a parameter in the coupled kicked rotor \cite{peres1984stability}, variation of kicking strength in the sawtooth map \cite{benenti2002quantum}, variation of the strength of potential \cite{Haug2005} and detuning of a trapping laser \cite{andersen2006}. These studies showed that the fidelity decays either exponentially or with some power law and may saturate to a value very close to zero after some relevant timescale that depends on the system. Devising methods for enhancing the fidelity of quantum systems over long periods of time is of critical importance for quantum control and quantum computing \cite{miquel1996factoring,gea1998qubit,benenti2001efficient,frahm2004quantum}. Dynamical decoupling \cite{viola1999dynamical}, quantum error correction \cite{bennett1996mixed,knill2000theory}, decoherence-free subspaces \cite{lidar1998decoherence} and the quantum Zeno effect  \cite{itano1990quantum,maniscalco2008protecting,Layden2015} are some of the methods proposed to enhance the stability of quantum computing.

In this Letter, we propose a method to stabilize systems against perturbations caused by variations in the external control parameter of a system. Our method ensures a finite lower bound on the fidelity decay and we show that this bound can be significantly greater than zero. The approach is generally applicable to any quantum system and experimentally simple to implement. Our method is based on a quantum description of the control parameters of the system. Consider a quantum system with Hamiltonian, $\lambda H_{\text{sys}}$, where $\lambda$ is an external control parameter that can be varied in some range. Now let us consider that the control parameter is quantum in nature: suppose we have an environment with Hamiltonian $H_{\text{env}}$ with eigenvalues, $\lambda$, and eigenstates, $\ket{\phi_{\lambda}}$. Then the joint state $|\phi_{\lambda} \rangle \otimes |\psi_{\text{sys}}\rangle$ evolves according to the Hamiltonian, $H_{\text{env}} \otimes H_{\text{sys}}$. If the initial state of the `system+environment' is a product state and the state of the environment is an eigenstate, the system evolves with the control parameter value $\lambda$ given by $\langle H_{\text{env}} \rangle$. Given this picture, we now consider a finite-dimensional  environment where the initial state of the environment is  a superposition of eigenstates rather than a single eigenstate. In such a case, we refer to $\langle H_{\text{env}} \rangle$ as the effective control parameter for the system. Its value depends on the initial state of the environment. We then study the stability of the system's evolution (as measured by the fidelity) with respect to changes in the effective control parameter of the system via changes in the initial state of the environment. We find that by appropriately choosing the initial superposition of states of the environment so that we introduce some degree of uncertainty in the control parameter, the system, perhaps surprisingly, can become significantly robust against changes in the effective control parameter. We quantify this stability by deriving a lower bound on the fidelity function and finding the maximum of this bound. We will see that this is applicable to any system, including highly fragile chaotic systems. We illustrate the method in the model of the quantum kicked top, which exhibits both regular and chaotic behavior .

\textit{System coupled to a single-qubit environment:} The quantum fidelity $F$ is defined to be the overlap of two states evolved from the same initial state: the first state evolves with the Hamiltonian, $H_0$, and the second evolves with a perturbed Hamiltonian, $H_{\epsilon} = H_0 + \epsilon V$, 
\begin{equation}
F(|\psi(t) \rangle, |\psi_{\epsilon}(t)\rangle) =  |\langle \psi(0)|U^{\dagger}_{\text{sys}}(t)U^{\epsilon}_{\text{sys}}(t)|\psi(0)\rangle|.
\label{sec2eq2}
\end{equation}
Here $U_{\text{sys}}$ and $U^{\epsilon}_{\text{sys}}$ are the time evolution unitary operators corresponding to the the unperturbed and the perturbed Hamiltonian respectively. 

 We first consider a system coupled to an environment consisting of a single qubit governed by the Hamiltonian, $H_{\text{qubit}} = \alpha \sigma_z + \beta \mathbf{1}$, where $\sigma_z$ is the Pauli operator. The unitary operator at time $t$ for the `system+qubit' is
\begin{equation}
U_{\text{total}}(t) =  \mathcal{T} \exp{\left(-\ii(\alpha \sigma_z + \beta \mathbf{1})\otimes \int_{0}^{t} H_{\text{sys}} dt \right)}.
\end{equation} 
Let $\rho(0)  = \rho_{\text{qubit}}(0)\otimes \rho_{\text{sys}}(0)$ be the initial state of the `system+qubit', where 
\begin{equation}
\rho_{\text{qubit}}(0) = \begin{bmatrix}
a & b \\b^* & 1-a
\end{bmatrix}.
\label{qubit1}
\end{equation}

The state of the total system (system $+$ qubit) after time $t$ is
\begin{eqnarray}
\rho(t) & = & U_{\text{total}}(t) \rho(0) U^{\dagger}_{\text{total}}(t) \notag \\
& = & a|0\rangle \langle 0| \otimes V_{\alpha}(t) \rho_{\text{sys}}(0) V_{-\alpha}(t) \notag \\ && + (1-a)|1\rangle \langle 1| \otimes V_{-\alpha}(t) \rho_{\text{sys}}(0) V_{-\alpha}(t) \notag \\ && + b|0\rangle \langle 1| \otimes V_{\alpha}(t) \rho_{\text{sys}}(0) V_{-\alpha}(t) \notag \\ && + b^*|1\rangle \langle 0| \otimes V_{-\alpha}(t) \rho_{\text{sys}}(0) V_{\alpha}(t)
\label{fidel24}
\end{eqnarray}
where 
\begin{eqnarray}
V_{\alpha}(t) & = & \mathcal{T} \exp{\left(-\ii \left( \beta + \alpha \right) \int_{0}^{t} H_{\text{sys}} dt \right)}, \label{fidel24b} \\
V_{-\alpha}(t) & = &\mathcal{T} \exp{\left(-\ii \left(\beta - \alpha \right) \int_{0}^{t} H_{\text{sys}} dt \right)}. \label{fidel24c}
\end{eqnarray}
Thus, the reduced state of the system at time $t$ after tracing out the environment is 
\begin{eqnarray}
\rho_{\text{sys}}(t) & = & \tr{_{\text{qubit}}(\rho(t))}  \nonumber \\
 & = & a \sigma_{\alpha}(t)  + (1-a) \sigma_{- \alpha}(t), 
\label{fidel3}
\end{eqnarray}
where the following notation has been used:
\begin{eqnarray}
\sigma_{ \alpha}(t) & = & V_{\alpha}(t) \rho_{\text{sys}}(0)  {V^{\dagger}_{\alpha}}(t), \nonumber \\
 \sigma_{- \alpha}(t) & = & V_{- \alpha}(t) \rho_{\text{sys}}(0) {V^{\dagger}_{- \alpha}}(t). 
 \label{fidel3b}
\end{eqnarray}
If we take a different initial state of the qubit
\[
\rho'_{\text{qubit}}(0) = \begin{bmatrix}
a' & b' \\b'^* & 1-a'
\end{bmatrix},
\label{qubit2}
\]
then if everything else remains the same as above, the state of the system at time $t$ will be
\begin{eqnarray}
\rho'_{\text{sys}}(t) = & a' \sigma_{\alpha}(t)  + (1-a') \sigma_{- \alpha}(t). 
\end{eqnarray}

The fidelity between $\rho_{\text{sys}}(t)$ and $\rho'_{\text{sys}}(t)$ is
\begin{eqnarray}
F(\rho_{\text{sys}}(t), \rho'_{\text{sys}}(t)) &=& F(a \sigma_{\alpha}(t) + (1-a) \sigma_{- \alpha}(t), a' \sigma_{\alpha}(t) \notag \\ && + (1-a') \sigma_{- \alpha}(t) ) \notag \\ & \geq & F(a \sigma_{ \alpha}(t),a' \sigma_{\alpha}(t)) \notag \\ && + F((1-a) \sigma_{-\alpha}(t), (1-a') \sigma_{-\alpha}(t) ) \nonumber \\
 & = & \sqrt{aa'} + \sqrt{(1-a)(1-a')},
\label{fidel1}
\end{eqnarray}
which is obtained using the concavity of the fidelity function for positive definite matrices, and the definition of the fidelity function for mixed states \mbox{$F(\rho_1,\rho_2) = \tr{\sqrt{\sqrt{\rho_1}\rho_2 \sqrt{\rho_1}}}$}.

Now, for $\rho_{\text{qubit}}$ and $\rho'_{\text{qubit}}$ to be valid quantum states,
\begin{eqnarray}
0 \leq a,a' \leq  1 & \\
\Rightarrow \sqrt{aa'} + \sqrt{(1-a)(1-a')} \geq & \text{ min} \{ \sqrt{a},\sqrt{1-a} \}.\notag
\label{condition1}
\end{eqnarray}

Thus, we arrive at the following lower bound on fidelity
\begin{equation}
F(\rho_{\text{sys}}(t), \rho'_{\text{sys}}(t)) \geq \text{min} \{ \sqrt{a},\sqrt{1-a} \}
\label{fidel21b}
\end{equation}
 for any $\rho_{\text{qubit}}(0)$ coupled to a qubit with initial state given by Eq. \eqref{qubit1}. By choosing a suitable initial state of the qubit, the lower bound of the fidelity can be raised to a maximum possible value of $1/\sqrt{2} \approx 0.7071$. This corresponds to the qubit's degree of freedom coupled to the system being in a state of maximal uncertainty (e.g., equal superposition state $1/\sqrt{2}(|0\rangle + |1\rangle)$, or a maximally mixed state).
 
 The effective value of the control parameter is $\langle \beta_{\text{eff}}\rangle = \langle H_{\text{env}}\rangle = \langle \alpha \sigma_z + \beta \mathbf{1} \rangle = \beta + \alpha(2a-1)$, where the value of $a$ depends on the state of the qubit. However, this $\beta_{\text{eff}}$ has a standard deviation, 
  \begin{equation}
 \begin{split}
     \text{std}(\beta_{\text{eff}}) & = \sqrt{\langle (\alpha \sigma_z + \beta \mathbf{1})^2 \rangle - \langle \alpha \sigma_z + \beta \mathbf{1} \rangle ^2} \\
     & = 2 \alpha\sqrt{a(1-a)}
 \end{split}
 \label{std}
 \end{equation}
 Thus, we obtain robustness in the fidelity at the cost of precision in the control parameter value.

\textit{System coupled to a harmonic oscillator environment:} We obtain a similar lower bound on the fidelity also for higher-dimensional environments, such as a harmonic oscillator, with the Hamiltonian, $\left(\beta \mathbf{1} + \alpha (a^{\dagger}a + \frac{1}{2}) \right)$. We will restrict the state of the environment to a finite-dimensional subspace, $|\psi_{\text{osc}}\rangle = \sum\limits_{k=0}^{l-1} c_k |k\rangle$. Here, $|k\rangle$ refers to the energy eigenstates of the oscillator. We follow exactly the same analysis here as for the qubit environment, except that the initial state of the `system + oscillator' is to be a pure state (for more tractable calculations).

Let $|\Psi(0) \rangle$ and $|\Psi'(0) \rangle$ be two initial states of the system coupled to the harmonic oscillator where the difference is only in the oscillator state,
\begin{eqnarray}
|\Psi(0) \rangle = (\sum_{k=0}^{l-1} c_k |k \rangle)\otimes |\psi_{\text{sys}}(0) \rangle, \\
|\Psi'(0) \rangle = (\sum_{k=0}^{l-1} c'_k |k \rangle)\otimes |\psi_{\text{sys}}(0) \rangle. 
\label{fidelity3}
\end{eqnarray}
We evolve these two states with $H_{\text{osc}}\otimes H_{\text{sys}}$ for time $t$. Let $\varrho_{\text{sys}}(t)$ and $\varrho'_{\text{sys}}(t)$ denote the reduced states of the system at time $t$. We compute the fidelity between these reduced states to be 
\begin{equation}
F(\varrho_{\text{sys}}(t), \varrho'_{\text{sys}}(t))   =  \sum_{k=0}^{l-1} |c_k||c'_k|   \geq  \underset{k \in \{0,...,l-1\} }{\text{min}} |c_k|.
\label{fidelity13}
\end{equation}
Thus, by choosing an appropriate initial state for the environment, the lower bound on fidelity can be raised to $1/\sqrt{l}$ (corresponding to an equal superposition state of `$l$' eigenstates), with $l=2$ being optimal. The effective value of the control parameter, $\beta_{\text{eff}}$, in this case, is $\langle \alpha (a^{\dagger}a + \frac{1}{2}) + \beta \mathbf{1} \rangle = \beta + \alpha(\sum\limits_{k=0}^{l-1} k |c_k|^2 + \frac{1}{2})$.

\textit{Stabilizing quantum chaos: the quantum kicked top.} We illustrate our method using an example - the quantum kicked top (QKT) \cite{haake1987classical}.
This is a time-periodic system governed by the Hamiltonian :
\begin{equation}
H = \hbar \frac{\beta}{2 j \tau} J_z^2 + \hbar p J_y \sum_{n= - \infty}^{\infty} \delta (t - n \tau),
\label{top1}
\end{equation}
where $J_y$ and $J_z$ are angular momentum operators, and $\beta$ and $p$ are external control parameters. $j$ is a constant of motion. The unitary operator for a time period `$\tau$' is given by $U=\exp{\left(-i \frac{\beta}{2j \tau}J_z^2 \right)} \exp{\left(-ipJ_y \right)}$. The QKT is an experimentally realized model \cite{chaudhury2009nature,neill2016ergodic} that exhibits regular as well as chaotic behavior upon variation of the control parameters, $\beta$ and $p$. It is thus a standard paradigm for studying quantum chaos both theoretically and in experiments {\color{red} }. 

\begin{figure}
%\centering{\includegraphics[width=0.5\textwidth]{plot1_without_env.pdf}}
\subfloat[]{\includegraphics[width=0.25\textwidth]{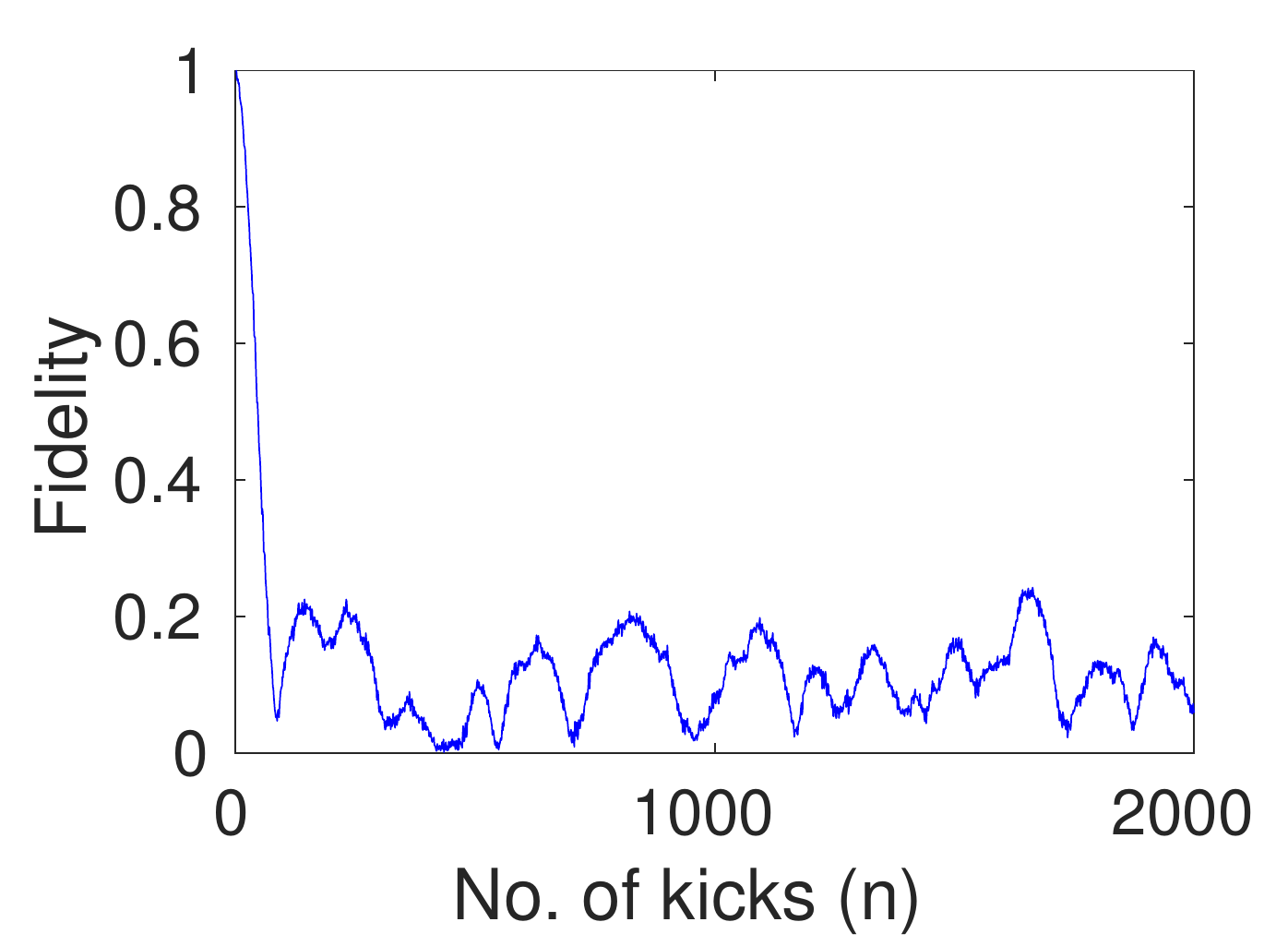}}
\subfloat[]{\includegraphics[width=0.25\textwidth]{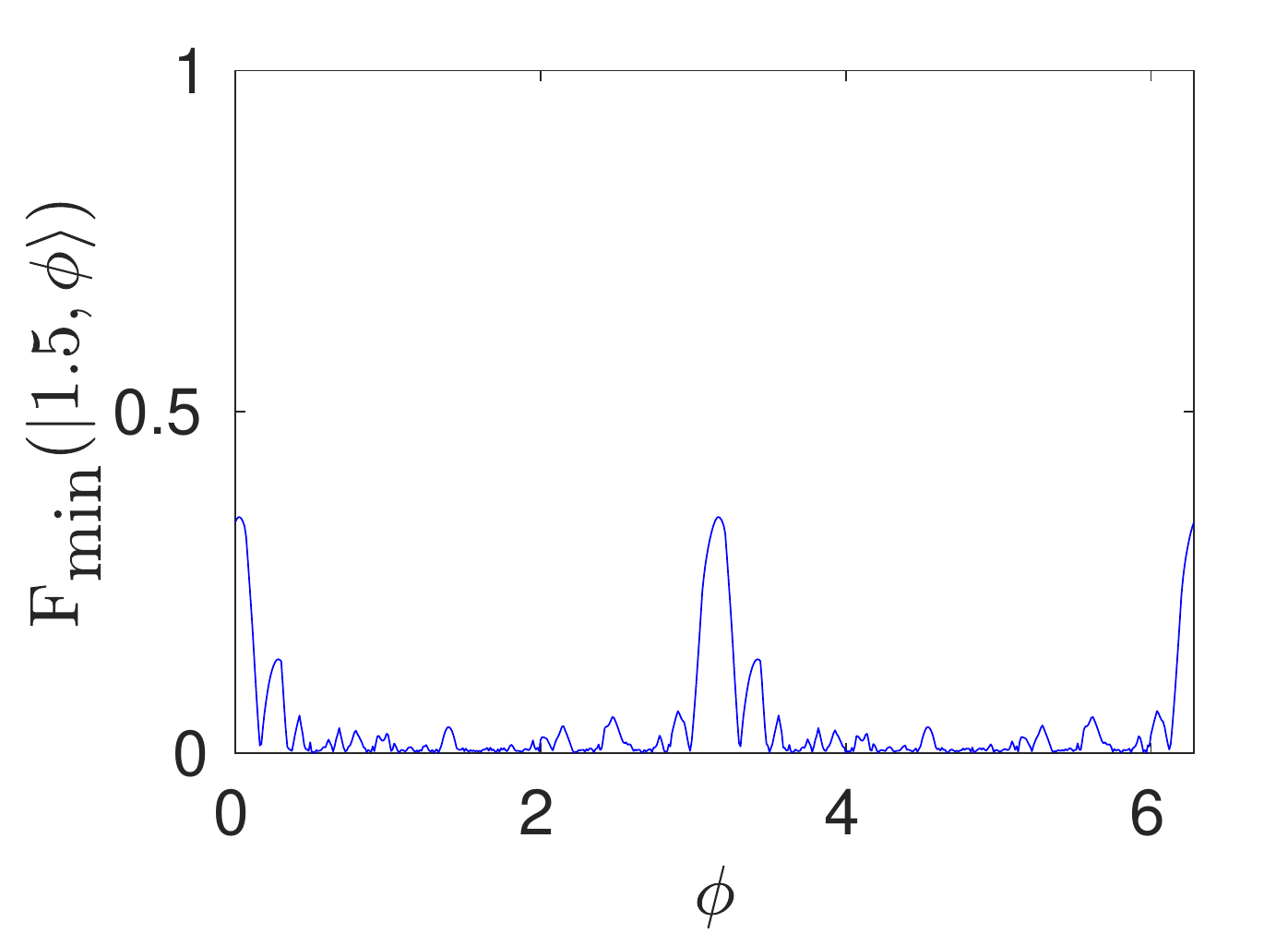}}
\caption{(a) Evolution of fidelity of QKT as a function of no. of time periods, $\vert \psi(0) \rangle = \vert 2.5,1.0 \rangle$. (b) Minimum fidelity over 1000 time periods as a function of $\phi$ for initial SCS states $|\theta,\phi \rangle$ with fixed $\theta=1.5$. Other parameter values are  $j = 100$, $\beta = 3.0$, perturbation strength $\epsilon = 0.01$, $p=\pi/2$ and $\tau = 1 $ }
\label{fig1}
\end{figure}

First, we show the sensitivity to perturbations in the control parameter of the QKT by numerically calculating the evolution of fidelity as given in Eq. \eqref{sec2eq2}. The initial state of the system is taken to be a spin coherent state (SCS), $\vert \psi (0) \rangle = \vert \Theta, \phi \rangle  =  \exp{ \left( \ii \Theta \left(J_x \sin{\phi})-J_y \cos{\phi} \right) \right) } \vert j,j \rangle $. This is a minimum uncertainty state such that the vector $(\langle J_x \rangle, \langle J_y \rangle, \langle J_z \rangle)$ points along the direction  $(\theta, \phi)$ on a sphere. As shown in Fig. 1, the fidelity reduces significantly in a few hundreds of time periods (referred to as no. of kicks) for a perturbation of $\epsilon = 0.01$ in the value of the control parameter $\beta$. To further demonstrate the sensitivity of the evolution of the QKT to the value of $\beta$, we plot the minimum fidelity over 1000 kicks for a range of initial SCS states $|\theta,\phi \rangle$ where $\theta$ is kept fixed and $\phi$ is varied (Fig.\ref{fig1}). The minimum fidelity drops below 0.1 for most of the SCS states, demonstrating that the lower bound of fidelity is zero for perturbations in the control parameter value.

We now demonstrate that by coupling the QKT to a qubit, the lower bound of fidelity can be raised to $0.7071$ as derived in Eq. \eqref{fidel21b}. The total Hamiltonian and unitary operator for one time period is
\begin{eqnarray}
H_{\text{tot}} & = & \frac{1}{2j\tau}\left(\alpha \sigma_z + \beta \mathbf{1}\right)\otimes J_z^2 + pJ_y\otimes \mathbf{1} \sum_{n}\delta(t-n\tau), \notag \\
U_{\text{tot}} &=& \exp{\left(\frac{-i}{2j}\left(\alpha \sigma_z + \beta \mathbf{1}\right)\otimes J_z^2 \right)}\exp{\left(-ip \mathbf{1}\otimes J_y \right)}.
\label{top3}
\end{eqnarray}
Let $|\Psi(0) \rangle = (d|0\rangle + \sqrt{1-|d|^2}|1\rangle)\otimes |\psi_{\text{top}}(0) \rangle$ be an initial state of the kicked top coupled to the qubit, where, for simplicity, we have chosen pure states for the qubit. The state of the coupled system after a time period $n\tau$ is $|\Psi(n\tau) \rangle = U^n_{\text{total}} |\Psi(0) \rangle$. To compute the fidelity, we consider another initial state $|\Psi'(0) \rangle = (d'|0\rangle + \sqrt{1-|d'|^2}|1\rangle)\otimes |\psi_{\text{top}}(0) \rangle$. In terms of $d$, the inequality in Eq. \eqref{fidel21b} becomes :
\begin{equation}
F(\rho_{\text{top}}(n\tau),\rho'_{\text{top}}(n\tau)) \geq \text{min}\{ |d|,\sqrt{(1-|d|^2)} \}. 
\label{top6}
\end{equation}
where $\rho_{\text{top}}(n\tau)$ and $\rho'_{\text{top}}(n\tau)$ are the reduced states of the kicked top obtained by tracing out the qubit in $|\Psi(n\tau)\rangle$ and $|\Psi'(n\tau)\rangle$ respectively. The effective value of the control parameter is $\beta_{\text{eff}} = \langle \alpha \sigma_z + \beta \mathbf{1} \rangle = \beta + \alpha(2|d|^2-1)$, which is equal to $\beta$ for $d=1/\sqrt{2}$.

\begin{figure}
\centering{\includegraphics[scale=0.4]{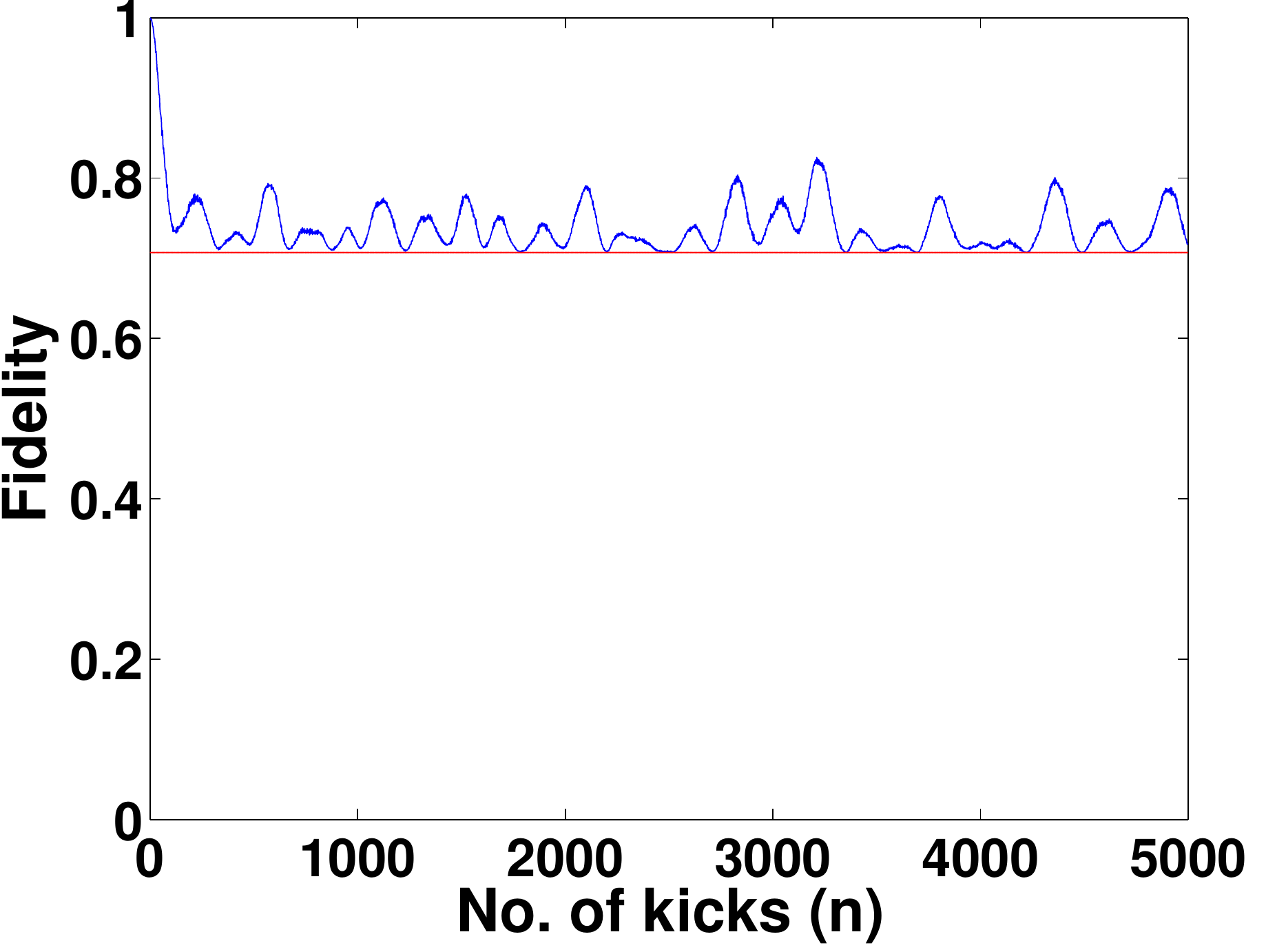}}
\caption{Evolution of fidelity of reduced state of QKT when it is coupled to the qubit. $j = 25$, $\vert \psi_{\text{top}}(0) \rangle = \vert 2,1.4 \rangle$, $\vert \psi_{\text{qubit}}(0) \rangle = \frac{1}{\sqrt{2}}(\vert 0\rangle + \vert 1\rangle), \vert \psi'_{\text{qubit}}(0) \rangle= \vert 0\rangle$, $\beta = 3.0$, $\alpha = 0.01$, $p=\pi/2$ and $\tau = 1 $}
\label{fig2}
\end{figure}

Fig. \ref{fig2}, shows that the fidelity evolution of the QKT coupled to a qubit respects the lower bound of Eq. \eqref{fidel21b}. We observe that $F(\rho_{\text{top}}(n\tau),\rho'_{\text{top}}(n\tau)) \geq \frac{1}{\sqrt{2}}$ for all $n$, where the initial state of the qubit in $|\Psi(0)\rangle$ and $|\Psi'(0)\rangle$ are $\frac{1}{\sqrt{2}}(\vert 0\rangle + \vert 1\rangle)$ and $\vert 0\rangle$ respectively. All the parameters pertaining to the kicked top are the same for Fig. \ref{fig1} and Fig. \ref{fig2} except $j$, and $\alpha = 0.01$ for Fig. \ref{fig2}. For the plot in Fig. \ref{fig2}, since $d=\frac{1}{\sqrt{2}}$, $\beta_{\text{eff}} = \beta = 3$, as illustrated previously. For this value of $\beta$, the kicked top exhibits chaotic dynamics in the classical limit. The standard deviation, $\text{std}(\beta_{\text{eff}}) = \alpha= 0.01 $, which is calculated using Eq. \eqref{std} and the parameters used for the plot. Thus, at a cost of uncertainty of $0.01$ in the value of the control parameter, we ensure that the fidelity always remains greater than $\frac{1}{\sqrt{2}}$. Although the perturbation of the control parameter in Fig. \ref{fig1} is the same as the standard deviation in the value of the effective control parameter value in Fig. \ref{fig2}, the evolution of the fidelity function for the two cases is very different. Coupling of the system to the qubit thus leads to robustness in fidelity. Furthermore, this robustness occurs even when the system under consideration exhibits chaotic behaviour. Our results are based on experimentally accessible parameters that can be implemented in experiments using currently existing technology \cite{chaudhury2009nature,Smith2013,Anderson2015}.

\textit{Discussion:} The lower bound of the fidelity function is  zero for the standard description of classical perturbations of the control parameter in the system Hamiltonian. Upon quantizing the control parameter using a discrete environment, the fidelity becomes lower bounded by a non-zero value, which can be maximized by appropriately choosing the initial state of the environment. In this quantized description, the old classical perturbation of the control parameter corresponds to the special case where the initial state of the environment is an eigenstate. By releasing this restriction of the environment to eigenstates and allowing superposition states, we have gained robustness in the fidelity decay.  The maximum of the lower bound on fidelity scales as $1/\sqrt{l}$, where $l$ is the dimension of the subspace, and thus tends to zero in the continuum limit.

Quantizing the control parameter in the proposed way yields a mixed-unitary non-Markovian quantum channel governing the evolution of the system. For example, for the qubit environment, the non-Markovian channel,
\begin{align}
    \Phi_t(\rho_{\text{sys}}(0)) =& a V_{\alpha}(t) \rho_{\text{sys}}(0) {V^{\dagger}_{\alpha}}(t)\\ \notag &+(1-a) V_{- \alpha}(t) \rho_{\text{sys}}(0) {V_{- \alpha}^{\dagger}}(t),\end{align} 
    governs the evolution of the state of the system, as evident from Eqs. \eqref{fidel3} and \eqref{fidel3b}. Non-Markovianity is a necessary condition to get the robustness in fidelity obtained in this paper. Let us consider a Markovian channel $\tilde{\Phi}_t(\cdot)$). Then
\begin{eqnarray}
\tilde{\Phi}_t(\cdot) & = & \tilde{\Phi}_{t/2}\circ \tilde{\Phi}_{t/2}(\cdot) \label{fidelity20a} \\
\Rightarrow \tilde{\Phi}_t(\rho) & = & a^2 V_{\alpha}^2 \left(t/2 \right) \rho {V^{{\dagger}^2}_{\alpha}} \left(t/2\right) \notag \\ && + a(1-a) V_{\alpha}(t/2)V_{- \alpha}(t/2)\rho {V_{- \alpha}^{\dagger}}(t/2) {V_{ \alpha}^{\dagger}}(t/2) \notag \\ && + a(1-a) V_{-\alpha}(t/2)V_{\alpha}(t/2)\rho {V_{ \alpha}^{\dagger}}(t/2) {V_{- \alpha}^{\dagger}}(t/2) \notag \\ && + (1-a)^2 V^2_{- \alpha}\left(t/2\right) \rho V_{- \alpha}^{{\dagger}^2} \left(t/2\right).
\label{fidelity20}
\end{eqnarray}
with an analogous expression for $\tilde{\Phi'}_t$ (which is the channel corresponding to a different initial state of the environment), with $a$ being replaced by $a^{\prime}$. Then, a  straightforward calculation yields
\begin{eqnarray}
F(\tilde{\Phi}_t(\rho),\tilde{\Phi}'_t(\rho)) & \geq & (\sqrt{aa'}+\sqrt{(1-a)(1-a')})^2. 
\end{eqnarray}
which is a smaller bound than the bound if it was non-Markovian $(\sqrt{aa'}+\sqrt{(1-a)(1-a')})$. In Eq. \eqref{fidelity20a}, since we have assumed the channel to be Markovian, we can just as well break the time $t$ into $n$ steps instead of 2 steps. Breaking it into $n$ steps will lead to a bound of $(\sqrt{aa'}+\sqrt{(1-a)(1-a')})^n$. Taking $n \rightarrow \infty$, the lower bound tends to zero. This shows that  a Markovian channel cannot achieve the lower bound on fidelity attained by the non-Markovian channel. This illustrates the importance of Markovianity.

Another interesting aspect is the limit $\alpha \rightarrow 0$ in the Hamiltonian of the environment, $H_{\text{env}}$. The lower bound obtained for fidelity is consistent for arbitrarily small $\alpha$. However, for $\alpha=0$, the fidelity becomes 1. This is because the channel corresponding to the two different initial states of the environment, $\Phi_t(\cdot)$ and $\Phi'_t(\cdot)$, turn out to be the same for the case of $\alpha=0$.  This apparent discontinuity can be understood by examining the time evolution of the fidelity decay. The initial rate of decrease of the fidelity function depends on $\alpha$, though the lower bound does not. The rate of decrease of the fidelity function gets smaller with decreasing $\alpha$. Thus, the fidelity function is continuous as a function of $\alpha$ for any time $t$ and we have pointwise convergence of the fidelity function. However, the fidelity function is not uniformly convergent $\alpha=0$, thus  the two limits, $\alpha \rightarrow 0$ and $t \rightarrow \infty$ do not commute with each other.

In summary, we know from the existing literature that for most quantum systems any small perturbation in control parameters can drive the system to very different states (in regular as well as chaotic systems), and the fidelity can drop very close to zero. We have shown that this drop in fidelity can be reduced and in fact bounded from below by considering the control parameter to be part of an environment system and by giving uncertainty to that control parameter. To see intuitively why this works, we may consider our system as a quantum device that measures a quantum observable of the environment, namely the control parameter. If the control parameter is prepared without uncertainty then our system may possess a pointer variable that measures the control parameter arbitrarily accurately. The fact that the pointer states overlap arbitrarily little implies that the fidelity can drop arbitrarily low. In contrast, if the control parameter possesses uncertainty, then so will any pointer variable of our system, indicating a lower bound on the fidelity. This lower bound then persists independent of the precise nature and strength of the interaction, except when the interaction strength is set to zero, which explains the discontinuity discussed above. In particular, contrary to expectations, the new results show that even chaotic systems need not be susceptible to fidelity decay. Since this mechanism for fidelity decay reduction does not depend on the specific details of the system and works well even if the environment is as small as a qubit, it should be of practical use.

\bibliography{references2}

\end{document}